\documentclass{article}
\usepackage{epsfig}

\def\be{\begin{equation}}
\def\ee{\end{equation}}
\def\bea{\begin{eqnarray}}
\def\eea{\end{eqnarray}}

\begin{document}

\begin{center}
{\LARGE \bf Dimensional Regularization and Mellin Summation in 
High-Temperature Calculations}
\\[4mm]
{\large D. J. Bedingham}
\\[4mm]
{\it Theoretical Physics, The Blackett Laboratory, Imperial College,\\ Prince Consort Road, 
London, SW7 2BW, U. K.\\ E-mail: dj.bedingham@ic.ac.uk}
\\[4mm]
\end{center}

\abstract{A general method for calculating asymptotic expansions of 
infinite sums in thermal field theory is presented. It is shown that 
the Mellin summation method works elegantly with dimensional 
regularization. A general result is derived for a class of one-loop
Feynman diagrams at finite-temperature.}

\section{Method}
The infinite sums often encountered in thermal Feynman diagrams are
commonly computed using the function $coth$, or one with similar properties,
to generate poles in
the complex plane whose residues correspond to the terms in the sum
\cite{kapusta}.
This transforms the summation into a contour integration and conveniently 
splits the zero-temperature and thermal contributions. 
The method ceases to be ideal when calculating the high temperature 
asymptotic expansion of such sums.
Cancellations occur between the thermal 
and non-thermal parts suggesting that the calculation could be streamlined. 
Here we propose a more concise method using dimensional 
regularization and the Mellin transform pair.
The sums we shall consider occur in one-loop calculations and
though these are well understood, the aim of this work is to outline a
convenient and general method for their evaluation in the high temperature 
limit. 

We recall first that the Mellin transform pair \cite{davis}  
can be written in the form
\bea {\cal M}[f;s]&=&\int_0^{\infty} x^{s-1} f(x) dx,\label{eq:mell}  \\
    f(x) &=& \frac{1}{2 \pi i} \int_{c-i\infty}^{c+i\infty}
        x^{-s} {\cal M}[f;s] ds. \label{eq:inv}
\eea
The transform normally exists only in a strip $\alpha < {\rm Re}[s] <
\beta$, and the inversion contour must lie in this strip:
$\alpha < c < \beta$.

We will find that representing a function as in (\ref{eq:inv})
is particularly useful for asymptotic analysis.
We will look at infinite sums of the type
\be I = \frac{1}{\beta}\sum_{n \neq 0}f(\omega_n) 
\ee
where
\be f(\omega_n) = \int\frac{d^d {\bf p}}{(2\pi)^d}\frac{\omega_n^{2t}}
	{({\bf p}^2+\omega_n^2+m^2)^\sigma}
\label{eq:f}
\ee
and $\omega_n=2\pi n/\beta$ for bosons (the method easily generalises to
fermionic integrals). We exclude the $n=0$ term of bosonic sums, 
however, as this may always be calculated independently.

By taking the Mellin transform of $I$ with respect to $\omega_n$, 
the sum can be represented using equations (\ref{eq:mell}) and 
(\ref{eq:inv}) as
\bea I &=& \frac{1}{2 \pi i \beta} \sum_{n\neq 0}\int_{c-i\infty}^{c+i\infty}
    \left(\frac{2\pi n}{\beta}\right)^{-s}{\cal M}[f;s] ds\\	
&=& \frac{1}{\pi i \beta} \int_{c-i\infty}^{c+i\infty}
    \left(\frac{2\pi }{\beta}\right)^{-s}\zeta(s)
	{\cal M}[f;s] ds.\label{eq:rep}
\eea
Provided that $c>1$, the interchange of the sum and integral is permitted
by the uniform convergence of the sum with respect to $s$. The Mellin 
transform is not necessarily convergent for $s>1$ and may require 
regulation. For example, one can use dimensional regularization and 
let $d$ in (\ref{eq:f}) become small.

We now take a closer look at the Mellin transform of $f$. 
From the definition of 
a $d$-dimensional integral \cite{collins}, we may write
\bea {\cal M}[f;s]&=&\int_0^{\infty}dy y^{s-1}\int
	\frac{d^d {\bf p}}{(2\pi)^d}
	\frac{y^{2t}}{({\bf p}^2+y^2+m^2)^\sigma}\\
&=& \frac{\Gamma(s/2+t)}{2\pi^{s/2+t}}\int d^{s+2t}{\bf y}
	\int\frac{d^d {\bf p}}{(2\pi)^d}
	\frac{1}{({\bf p}^2+{\bf y}^2+m^2)^\sigma}.
\eea
We may further combine these integrations to give
\be {\cal M}[f;s] = \frac{\Gamma(s/2+t)}{2\pi^{s/2+t}}
	\int\frac{d^{d+s+2t} {\bf p}}{(2\pi)^d}
	\frac{1}{({\bf p}^2+m^2)^\sigma}.
\label{eq:int}
\ee

We now see why the Mellin summation technique is particularly suited
to this type of calculation. The Mellin transform neatly combines 
with the $d$-dimensional integral leaving a $(d+s+2t)$-dimensional 
integral which can easily be evaluated in terms of $\Gamma$-functions:
\be {\cal M}[f;s] = \frac{\Gamma(s/2+t)}{2(4\pi)^{d/2} }
	\frac{\Gamma (\sigma-d/2-s/2-t)}
	{ m^{2\sigma-d-s-2t} \Gamma(\sigma)}.
\label{eq:gam}
\ee

With $t,\sigma>0$, the integral is convergent when $d<2(\sigma-t)-s$
and so we must choose $2(\sigma-t)-d>c>1$. This can always be 
achieved by dimensional regularization.

Combining equations (\ref{eq:rep}) and (\ref{eq:gam}) we have
\bea I&=& \frac{m^{-2\sigma+d+2t}}{\Gamma(\sigma)(4\pi)^{d/2}\beta}
	\frac{1}{2\pi i }\times\nonumber\\
	&\times & \int_{c-i\infty}^{c+i\infty}
    \left(\frac{2\pi }{m\beta}\right)^{-s}\zeta(s)
     \Gamma(s/2+t)\Gamma (\sigma-d/2-s/2-t) ds.
\eea
From the asymptotic behaviour of the integrand we conclude that
we may close the integration path at infinity in the positive real 
half-plane. The only poles which are enclosed within the contour
are those due to $\Gamma (\sigma-d/2-s/2-t)$. This has simple 
poles at $s=2(\sigma-t)-d+2k$ for $k=0,1,2\ldots\infty$.

Evaluating the residues we find
\bea I&=&\frac{2(4\pi)^{-d/2}}{\Gamma(\sigma)\beta}
	\left(\frac{2\pi}{\beta}\right)^{d-2(\sigma-t)}
	\times\nonumber\\
	&&\times\sum_{k=0}^{\infty}\frac{(-1)^k}{k!}
	\zeta\left(2(k+\sigma-t)-d\right)\Gamma(k+\sigma-d/2)
	\left(\frac{m\beta}{2\pi}\right)^{2k}
\label{eq:result}
\eea  
which is our general result.

\section{Example: Scalar Tag Diagram}
Consider as a specific example the sum
\be J=\frac{\mu^{2\epsilon}}{\beta}\sum_{n\neq 0}\int
    \frac{d^{3-2\epsilon}{\bf q}}{(2\pi)^{3-2\epsilon}}
    \frac{1}{({\bf q}^2 +\omega_n^2 + m^2)}
\ee
which is the tag diagram of scalar field theory. 
We may use equation (\ref{eq:result}) with $\sigma=1,t=0$ and 
$d=3-2\epsilon$, to obtain 
\bea J&=&\frac{1}{(4\pi)^{1/2}\beta^2}
    \left(\frac{\mu^2\beta^2}{\pi}\right)^{\epsilon}
    \times\nonumber\\ &&\times\sum_{k=0}^{\infty}
    \frac{(-1)^k}{k!}\zeta(2k-1+2\epsilon)\Gamma(k-1/2+\epsilon)
    \left(\frac{m\beta}{2\pi}\right)^{2k}.
\eea
As we let $\epsilon$ tend to zero all terms in this series are well
defined except for the $k=1$ term. This term has a pole in the 
$\zeta$-function corresponding to the zero-temperature divergence.
For small $\epsilon$ we have
\bea  \zeta(1+2\epsilon)&=&\frac{1}{2\epsilon}+\gamma+{\cal O}(\epsilon)\\
    \Gamma(1/2+\epsilon)&=&\Gamma(1/2)\left[1-\epsilon(2\ln2+\gamma)
    +{\cal O}(\epsilon^2)\right]\\
    \left(\frac{\mu^2\beta^2}{\pi}\right)^{\epsilon}
    &=& 1 + \epsilon\ln\frac{\mu^2\beta^2}{\pi}
    +{\cal O}(\epsilon^2)
\eea
where $\gamma$ is the Euler-Mascheroni constant, resulting in
\be J_{k=1} = \frac{m^2}{(4\pi)^2}\left[\left(\frac{1}{\epsilon}
    -\gamma+\ln 4\pi\right)+2\ln\frac{\mu\beta e^{\gamma}}{4\pi}
    \right] + {\cal O}(\epsilon).
\ee
The first few terms in the series are then given by
\bea J&=&\frac{1}{12\beta^2}+
    \frac{m^2}{(4\pi)^2}\left[\left(\frac{1}{\epsilon}
    -\gamma+\ln 4\pi\right)+2\ln\frac{\mu\beta e^{\gamma}}{4\pi}
    \right]\nonumber\\
   && +\frac{2(m^2)^2\beta^2\zeta(3)}{(4\pi)^4}+ {\cal O}(\beta^4).
\eea

\section{Summary}
\begin{figure}[t]
\epsfxsize=10pc 
\[\epsfbox{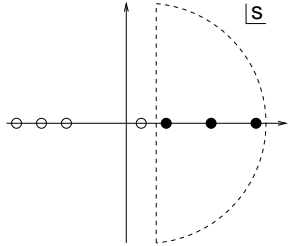}\] 
\caption{The Mellin inversion contour is closed in the positive
    real half plane. The only poles enclosed by this contour
    are the UV divergences of the Mellin transform (filled dots).
	 Poles external to the contour
    include that of the $\zeta$-function at $s=1$ and those due
    to infrared divergences in the Mellin transform.}
\end{figure}

\begin{figure}[t]
\epsfxsize=10pc 
\[\epsfbox{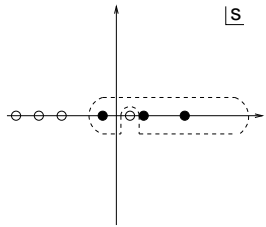}\] 
\caption{As the regulator is removed, the poles move to the left. A
    pole inside the contour may tend to coincide with
    $\zeta$-function pole - this will generate the zero-temperature
    divergence.}
\end{figure}

We may summarise the method in a prescription for generating the high 
temperature expansion of one-loop thermal Feynman diagrams.

(i) We express the sum as in equation (\ref{eq:rep}). The Mellin
transform must be regulated by suitable choice of dimension, $d$,
such that it exists for $1<{\rm Re}[s]$.
The inversion integration can then be defined with $1<c$.

(ii) The contour is closed in the right-hand half-plane (see figure 1). 
In order to know if this arc gives a contribution to the sum we should 
check the asymptotic value of the integrand of (\ref{eq:rep}) as
$|s|\rightarrow\infty$. 

(iii) We then deform the contour as we allow our regulator $d$ to
tend to its original value (figure 2). The result is the same if
we first calculate the residues and then reinstate $d$.

We may anticipate divergences resulting from pinch singularities between
poles inside and outside the contour. These will correspond to the
zero-temperature divergences and are also responsible 
for generating the $\ln\mu\beta$ terms as seen in the example
($\ln\mu/m $ terms do not appear at high temperature).
In general, it is seen that contributions 
to the sum result from poles in $s$ of the $(d+s+2t)$-dimensional integral
of equation (\ref{eq:int}).

\section*{Acknowledgements}
I am grateful to Tim Evans for his input of ideas and guidance.

\end{document}